# Spin-to-orbital angular momentum conversion in dielectric metasurfaces

Robert Charles Devlin[1]*, Antonio Ambrosio[1,2]*, Daniel Wintz[1], Stefano Luigi Oscurato[3], Alexander Yutong Zhu[1], Mohammadreza Khorasaninejad[1], Jaewon Oh[1,4], Pasqualino Maddalena[3], Federico Capasso[1]

[1]*Harvard John A. Paulson School of Engineering and Applied Sciences, Harvard University, Cambridge, Massachusetts 02138, USA*

[2]*CNR-SPIN U.O.S. Napoli, Complesso Universitario di Monte Sant'Angelo, Via Cintia, 80126 – Napoli, Italy*

[3]*Dipartimento di Fisica "E. Pancini", Università di Napoli Federico II, Complesso Universitario di Monte Sant'Angelo, Via Cintia, 80126 – Napoli, Italy*

[4]*University of Waterloo, Waterloo, ON N2L 3G1, Canada*

*These authors contributed equally to this work

*corresponding authors: antonio.ambrosio@cnr.it, capasso@seas.haravrd.edu*

**Spin-to-orbital-angular-momentum conversion has attracted considerable interest as a tool to create exotic light beams, leading to the emergence of novel devices that implement this function. These converters exploit the geometrical phase to create helical beams of handedness determined by the chirality of the incident light. This property is finding important applications in quantum optics thanks to the demonstration of liquid crystal spin-to-orbital angular momentum converters (SOC) known as *q*-plates. Here we demonstrate high-efficiency SOCs in the visible based on dielectric metasurfaces that generate vortex beams with high and even fractional topological charge and show for the first time the simultaneous generation of collinear helical beams with different and arbitrary orbital angular momentum. We foresee that this versatile method of creating vortex beams, which circumvents the limitations of *q*-plates, will significant impact microscopy and vector beam shaping.**

A helical mode of light is an optical field whose azimuthal phase evolution around the propagation axis (*z*) has the form $\exp[i\ell\varphi]$, $\varphi$ being the azimuthal angle and $\ell$ (an integer) called *topological charge* of the beam. The wavefront of a helical mode of charge $\ell$ is constituted by $|\ell|$ helical surfaces twisted together, whose handedness is set by the sign of $\ell$, resulting in a topological singularity (*optical vortex),* along the propagation axis [1]. Such vortex beams carry an average of $\ell\hbar$ orbital angular momentum (OAM) per photon [2,3]. On the other hand, circularly polarized helical modes also carry a spin angular momentum of $\pm\hbar$ per photon, depending on the polarization handedness. Such beams are central to the field of singular optics [4] and have found numerous applications such as optical trapping [5] where the angular momentum is a powerful manipulation tool to spin the trapped object [6,7] as well as to control its orientation [8].

The characteristic screw-type dislocation of a helical mode can be imposed on the wave-front of a propagating beam by means of different devices, for example, pitch-fork holograms [9,10] or cylindrical and axicons lenses and reflectors [11,12]. Additionally, helical modes can be also produced by exploiting the geometrical phase (also known as Pancharatnam-Berry (PB) phase) [13,14], to create inhomogeneous gratings for the wavefront reshaping [15,16]. In these spin-orbital angular momentum converters (SOC) the OAM of the vortex beam is entangled with the spin momentum of the illuminating light: switching the handedness of the illuminating beam polarization (spin momentum) flips the handedness of the vortex. Locking the OAM to the spin momentum has unique applications in quantum computing and communications, allowing encoding of quantum units [17] and fast switching related to the modulation of incident polarization of light [18].

More recently, the wavefront manipulation allowed by metasurfaces [19] has been used to produce a variety of PB optical elements, e.g., lenses [20,21] and spin-OAM converters in the near-infrared [15,16,22,23]. Similar approaches have allowed working with visible light although with low transmission efficiency in the bluest part of the spectrum [24,25,26, 27,28]. To date, the most versatile spin-orbital momentum converters for visible light are the liquid crystal devices developed by Marrucci *et al*. in 2006 and known as *q*-plates [29]. They have found numerous applications in quantum optics although limited by degradation effects, fabrication reproducibility and resolution in defining the extent of the topological singularity region [30,31,32,33,34].

In order to describe some general features of a SOC based on PB phase, it is useful to define the orientation angle $\alpha(x,y)$ of the optical axis (fast or slow) of each element of the device in the transverse plane (*x-y* plane). Regardless of the constituents, if each element imposes a π phase delay between the field transverse components, an incident uniform left-circularly polarized beam $E_{in} = E_0 \times [1,i]$ is turned into the beam $E_{out} = E_0 \exp[i2\alpha(x,y)] \times [1,-i]$ that is right-circularly polarized with a geometrical phase $2\alpha(x,y)$ in the transverse plane. Analogously to what reported in the first description of a *q*-plate [29], if the azimuthal variation of the angle $\alpha$ in the PB-device follows the relation $\alpha = q\varphi + \alpha_0$, the incident wave front is then turned into a helical wave front composed of $2|q|$ intertwined helical surfaces which carries an orbital angular momentum $\ell = \pm 2q$, where the sign depends on the handedness of the incident light polarization ($\alpha_0$ is a constant). For instance, if $q=1$ and the incident light is left-circularly polarized (spin momentum of $+\hbar$), the out coming light is right-circularly polarized (spin momentum of $-\hbar$) with an OAM per photon of $2\hbar$ and zero net angular momentum transferred to the device (Fig. 1 **a**). For $q \neq 1$ there will be a net angular momentum exchange with the PB-device to preserve the total angular momentum of the system.

In our devices, the constitutive elements (nanofins) are sub-wavelengths dielectric resonators [35,36,37,38] made of $TiO_2$ [39] (Supplementary Information). Each fin is 250nm long, 90nm wide and 600nm tall. The radial distance between two fins is of 325nm (Fig. 1 **b**). Figure 2 **a** and **b** show the scanning electron microscope (SEM) images of the devices with $q=0.5$ and $q=1$ ($|\ell|=1$ and $|\ell|=2$ respectively). The insets of Fig. 2 **a** and **b** show the devices as imaged in cross-polarization: the azimuthal variation of the optical field polarization direction after the device is turned into $4q$ intensity lobes by the second polarizer.

In order to fully characterize the vortex beams that our devices produce, a Mach-Zehnder interferometer was used as shown in Figure 2 **c**. In this configuration, the source beam (a solid state laser emitting at 532nm with power lower than 2mW) is split in two beams by means of a 50/50 beam splitter. Half of the light (upper arm of the interferometer) passes through a linear polarizer (LP1) and a quarter waveplate (QWP1) to produce a circularly polarized beam incident on the device. The vortex beam created by the device was then passed through a polarization filter made of a quarter waveplate (QWP2) and a linear polarizer (LP2) in cross-polarization with respect to QWP1 and LP1. This polarization filter is used to eliminate non-converted light passing through the device (Supplementary Information). The reference beam propagates in the lower arm of the interferometer and passes through a half waveplate (HWP) to acquire the same polarization of the helical mode in port 1. This maximizes the intensity modulation (thus the contrast) in the interference pattern.

Figure 2 **d** shows the intensity distribution of a vortex beam with $|\ell|=1$, generated by the device in Fig. 2 **a**, in a transverse plane (plane of the CCD camera at port 1 of the setup) at about 45cm from the device exit plane. Figure 2 **f** shows the intensity profile for the vortex beam with $|\ell|=2$ generated by the device of Figure 2 **b**. The four insets of Figure 2 **e** and **g** show the intensity patterns produced in the plane of the camera by interfering the vortex beam with the reference beam. Such interference experiments are widely used to reveal phase singularities [4]. The pitchfork-like interference was obtained when the vortex beam was interfered with a Gaussian beam incident at an angle (the incident angle sets the fringes spacing). Instead, when the vortex beam was collinear with the reference beam from the lower

arm of the interferometer, a spiral was obtained as interference picture, with a number of arms equal to the topological charge of the vortex beam. The handedness of the incident circularly polarized light sets the orientation of the pitchforks and spirals.

Figure 3 shows how our approach can be used to produce optical vortices with higher values of topological charge, $|\ell|=5$ (Fig. 3 **a-d**) and $|\ell|=10$ (Fig. 3 **e-h**). Each individual device is 500μm in diameter and all devices are on the same glass substrate of 1 inch diameter (Supplementary Information). This allows one to mount the device on standard opto-mechanical components and to select the desired topological charge just by translating the corresponding device into the laser beam path.

Another important feature of our devices is related to the localization of the beam singularity. The fabrication process is based on atomic layer deposition (ALD) and electron beam lithography (EBL) (Supplementary Information). This guarantees high resolution and reproducibility, resulting in precise definition of the singularity region and improving the vortex beam quality. For example, the $q=0.5$ device has a singularity region smaller than 3μm (Supplementary Information).

In our devices we reached absolute efficiencies (the amount of light from the illuminating beam that is actually converted into the helical mode) of 60% (Supplementary Information). This makes them usable for practical applications.

As further demonstration of the versatility of our approach, we designed a SOC that produces a vortex beam with fractional topological charge. This is possible when a non-integer phase discontinuity is introduced in the azimuthal evolution of the helical mode. In this case, Berry described the optical vortex as a combination of integer charge vortices with a singularity line in the transverse plane surrounded by alternating optical single charge vortices [40,41]. From a quantum optics point of view, the average angular momentum per photon has a distribution peaked around the nearest integer value of the topological charge and a spread proportional to the fractional part of the charge [42]. We fabricated a SOC producing a 6.5 topological charge vortex beam. Figure 4 **a** shows the intensity distribution of the resulting helical mode at about 55μm from the device plane (Supplementary Information) and Fig. 4 **b** shows pitchfork-like interference obtained in the Mach-Zehnder configuration of Figure 2 **c**. The phase singularity line predicted for such vortices is evident. The interference pattern (Figure 4 **b**) also shows the line of alternating vortices (single line pitchforks) along the singularity line. For half odd-integer values of the OAM, two helical modes with same OAM but phase singularities lines with a relative π orientation are orthogonal [42]. This has been used, for instance, to observe high-dimensional photon entanglement [43,44].

Our approach to SOC enables a new and unique feature, the generation of collinear beams with different OAM, a functionality that cannot be achieved with liquid crystals. To demonstrate this concept, we designed an interlaced $q = 2.5$ and $q = 5$ device (Figure 5 **a)**. Two metasurfaces with different azimuthal patterns are interleaved by placing the nanofins at alternating radii. Although they have different topological charges ( $|\ell| = 5$ and $|\ell| = 10$ ), the beams emerge collinearly from the device, interfering transversely to the propagation direction. Figure 5 **b** and **c** show the intensity patterns recorded on transverse planes (far from the device) for opposite handedness of the incident light. It is evident that the two interference patterns are flipped according to what is expected for beams with opposite topological charges. Figure 5 **c** and **d** show the calculated interference patterns of two collinear helical modes of topological charges 5 and 10 with opposite handedness. These interference patterns are close to what we found experimentally if we assume for the charge 5 beam a Rayleigh range three times greater than for the charge 10 beam. This seems reasonable since the interlaced structure has different geometrical parameters for the two topological charges, in particular different dimension of the singularity regions.

It is important to note that each nanofin in our device has two interfaces, glass-$TiO_2$ and air-$TiO_2$. Illuminating one side or the other, as in Figure 5 **f**, does not alter the phase delays imposed by the nanofins (Supplementary Information) but only slightly affects light coupling into the latter, due to the different reflectance of the air-$TiO_2$ and glass-$TiO_2$ interfaces. We measured a small decrease (< 5%) in the device efficiency when illuminating from the air-side, due to the lower refractive index.

In the setup of Figure 5 **f**, the beams illuminating the sample from opposite interfaces have opposite handedness (Methods). The *double-face* characteristic of our devices together with the illumination configuration of Figure 5 **f** allows one to simultaneously generate similar beams with opposite topological charges. This configuration was also used to obtain the intensity distributions of Figure 5 **b** and **c** representing the helical modes at optical ports 3 and 2 respectively.

Although we limited our interlaced design to two collinear beams, it is possible to produce three or more collinear vortices simultaneously as well as $+\ell$ and $-\ell$ collinear vortices (Supplementary Information). This can find important applications in entanglement and quantum computing experiments. Moreover, the quantum description of the photon statistics produced by our interlaced device has never been investigated and represents a stimulating direction for future work. Finally, although we did not test our devices at high incident power, we expect good tolerance to heating since $TiO_2$ has an intensity damage threshold of 0.5 J/cm$^2$ in the femtosecond regime [45]; thus we envision using such devices for non-linear optics with pulsed lasers.

In summary we have demonstrated that the interaction of light with designer metasurfaces can lead to the generation of complex wavefronts characterized by arbitrary integer and fractional topological charges and in particular to co-propagating beams with different orbital angular momenta. Our approach represents a major advance in design with respect to liquid crystals devices and as such has considerable potential in several areas of optics and photonics, ranging from quantum information processing to optical trapping and complex beam shaping.

## Methods

### Device fabrication

All devices used above were fabricated on a fused silica substrate. Resist was spun at 1750 rpm in order to achieve a thickness of 600 nm. The resist was then baked at 180 $^o$C for 5 mins. The patterns were exposed using electrom beam lithography (ELS-F125, Elionix Inc) and developed in o-Xylene for 60 s. For the ALD (Savannha, Cambridge Nanotech) of $TiO_2$, TDMAT precursor was used to avoid chlorine contamination and the system was left under continuous 20 sccm flow of $N_2$ carrier gas and maintained at 90 $^o$C throughout the process. Reactive ion etching was carried out on Unaxis ICP RIE with a mixture of $Cl_2$ and $BCl_3$ gas (3 and 8 sccm, respectively) at a pressure of 4 mTorr, substrate bias of 150 V and ICP Power of 400 W. The samples were finally placed in 2:1 sulfuric acid:hydrogen peroxide to remove any residual electron beam resist.

### Double-side illumination

The double-side illumination (figure 5 **f**) is simply obtained by rotating by 90° the beam-splitter at the exit of the Mach-Zehnder configuration of Figure 2 **c**. In this case, there are two light beams, whose power can be made equal by suitably balancing the two arms, circularly polarized with opposite handedness that simultaneously illuminate the device from opposite sides at normal incidence. In this configuration, the helical modes propagating towards optical port 2 and 3 have also opposite wave-front handedness.

## Acknowledgements

R.C.D is supported by a Draper Laboratory Fellowship. We also acknowledge financial support from Air Force Office of Scientific Research (AFOSR) contract

**Author Contributions**

A.A. conceived the idea and designed the devices. R.C.D. fabricated the devices. A.A. and R.C.D. conducted the experiments and analyzed the results with help from D.W., A.Y.Z. and S.L.O. on specific tasks of the project. M.K and J.O. optimized the nanofins by means of numerical simulations. A.A., R.C.D. and F.C. wrote the paper. All authors discussed the results and commented on the manuscript.

**Additional information**

Supplementary information is available in the online version of the paper.

**Competing financial interests**

The author declare no competing financial interests.

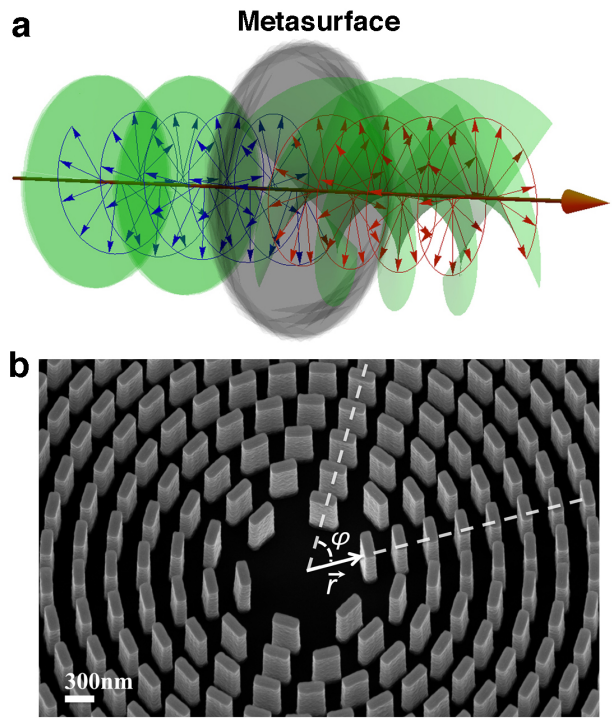

**Figure 1 | Nanofin-based spin-orbital momentum converter. a**, Schematic of the working principle of a spin-orbit converter. A left circularly polarized beam with plane wavefront is turned into a right circularly polarized helical mode. In this representation the helical mode has a topological charge equal to 2, as the wavefront is composed of two intertwined helices. **b**, Side-view SEM image of one of our devices ($q=1$) showing the orientation of the TiO$_2$ nanofins on the glass substrate, $\varphi$ is the azimuthal angle and |r| is the distance from the center.

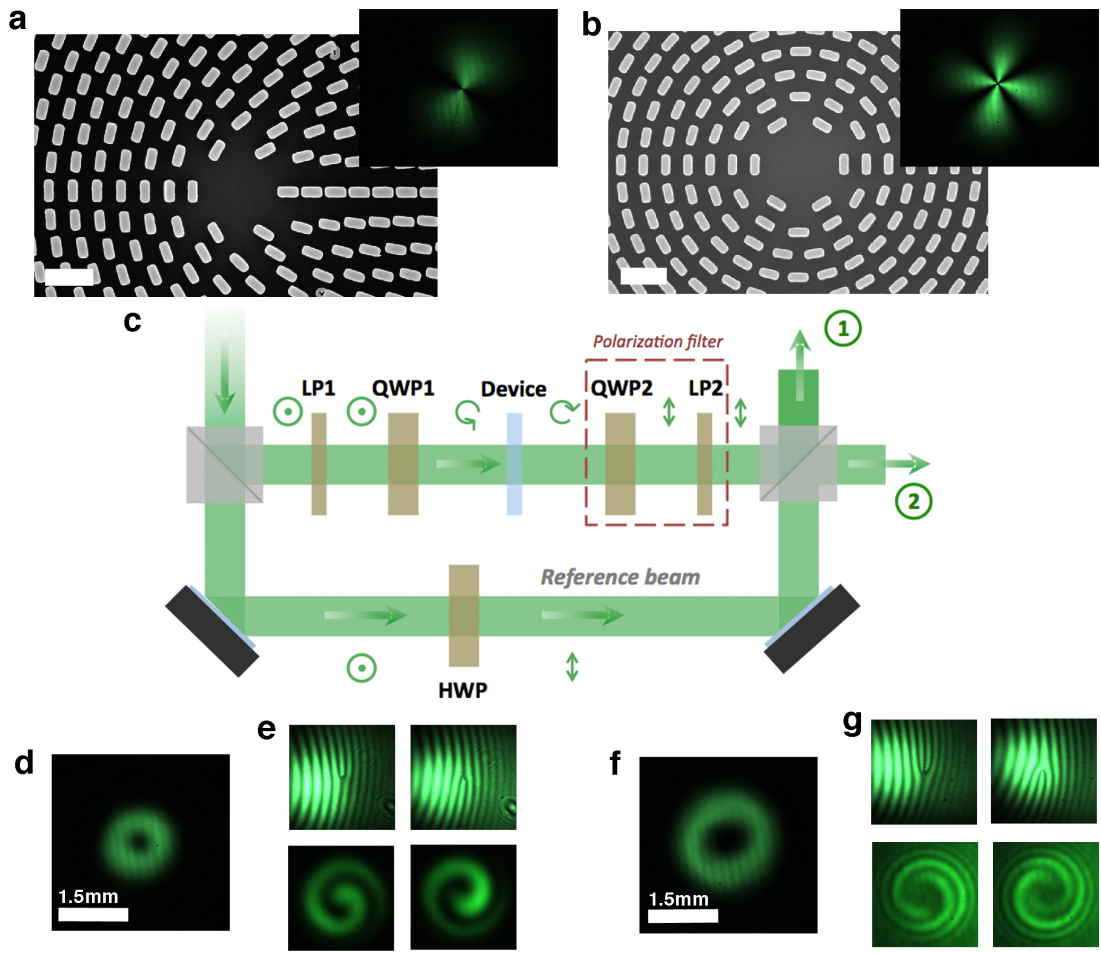

**Figure 2 | Devices characterization. a** and **b**, Scanning electron microscopy micrograph of $TiO_2$ -based spin orbit converters with $q = 0.5$ and $q = 1$ respectively (scale bar = 650nm). The insets show the devices observed in cross-polarization at the design wavelength of 532nm. **c**, Sketch (top view) of the interferometric setup used to characterized the devices. The interference of the helical mode and the reference beam was monitored at port 1 by means of a CCD. The polarization state of the beam after each optical element is sketched. The polarization after the first polarizer (LP1) is linear and perpendicular to the optical table. Light becomes circularly polarized after the first quarter waveplate (QWP1). The helical mode generated by the device is circularly polarized with opposite handedness. The helical mode after the polarization

filter (QWP2 followed by LP2) is linearly polarized parallel to the optical table. The reference beam in the lower arm of the interferometer becomes also linearly polarized parallel to the optical table after passing through a half waveplate (HWP) that rotates the polarization direction by 90°. **d**, Transverse intensity distribution of the vortex beam generated by the device of Fig. 2 **a**. This beam has a topological charge equal to 1. **e**, Interference patterns obtained with tilted reference beam (pitchforks) or collinear reference beam (spirals) in the setup of Figure 2 **c**. The flipped features result from opposite handedness of the beam that illuminates the device. **f**, **g**, Same as **d** and **e** for the device in Figure 2 **b**.

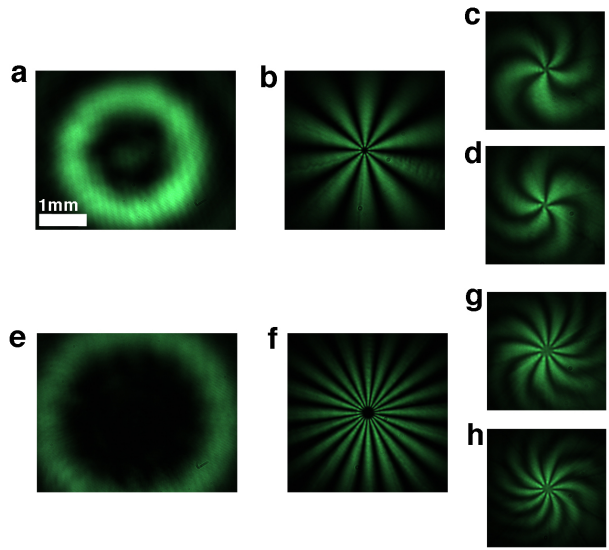

**Figure 3 | High-order helical modes. a**, Transverse intensity profile of a beam with topological charge 5 generated by means of our $q = 2.5$ spin-orbital momentum converter. **b**, The $q = 2.5$ device imaged in cross-polarization. **c**,**d** Images of the interference patterns obtained with a collinear reference beam (Figure 2 **c**) for incident left or right circular polarized light. **e**,**f**,**g**,**h**, Same as for **a**,**b**,**c**,**d** for the topological charge 10 beam and the $q = 5$ device. Scale for **e** is same as shown in **a**.

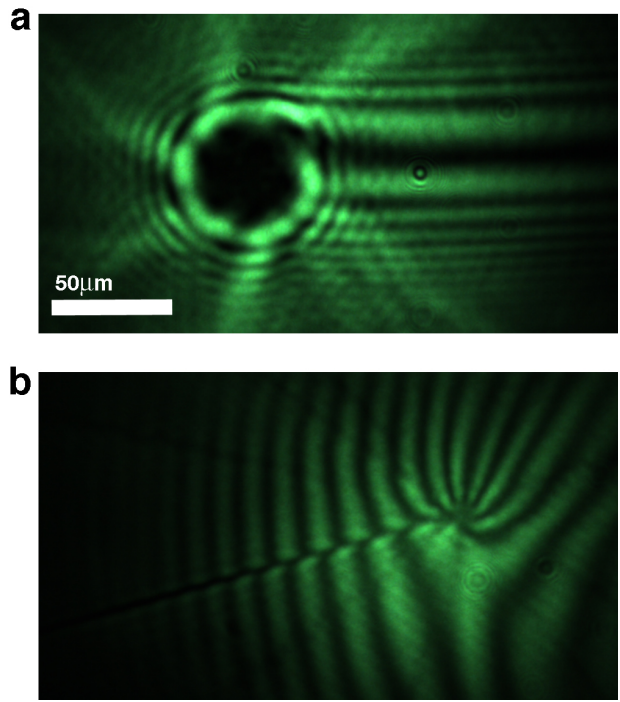

**Figure 4 | Fractional helical mode. a**, Transverse intensity distribution of a 6.5 topological charge beam at 55μm from the device exit plane. **b**, Interference pattern arising from the interference with a reference beam at oblique incidence. The resulting pitchfork pattern shows the singularity line surrounded by alternating single charge vortices, a characteristic feature of fractional helical modes. The direction of the singularity line in **a** and **b** is the same although in these figures they are on opposite directions due to the camera orientation during the experiment.

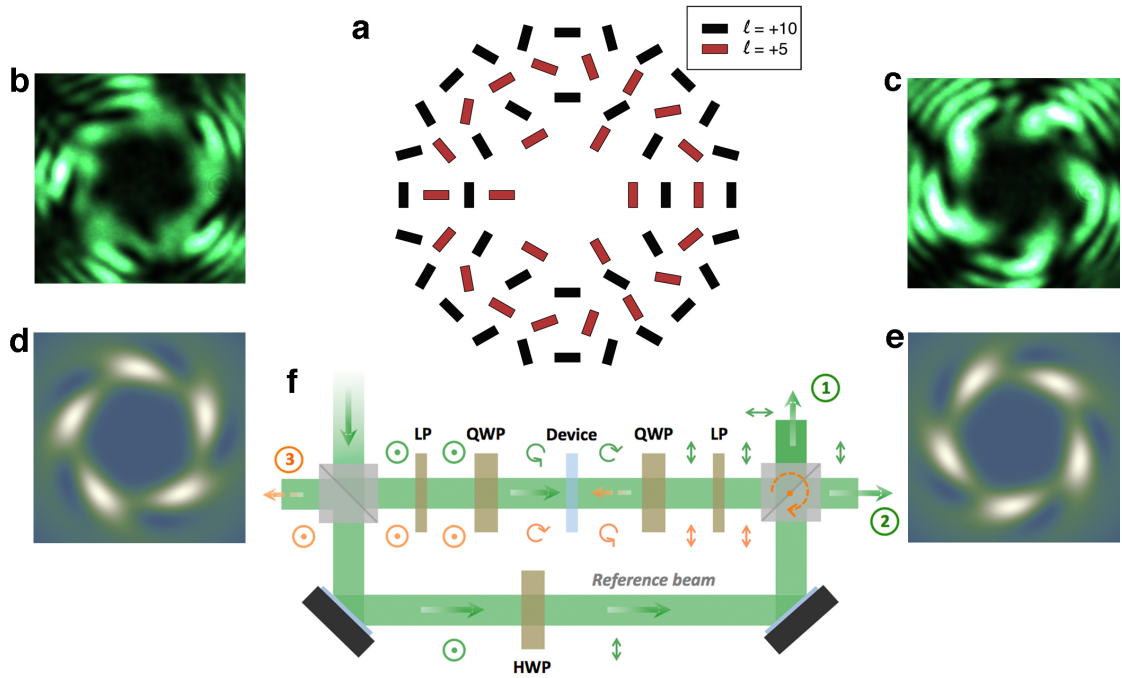

**Figure 5 | Generation of collinear helical beams of different topological charge. a**, Representation of the nanofins azimuthal distribution in the inner part of metasurface device with interleaved patterns that generates collinear beams with topological charges $|\ell|=5$ and $|\ell|=10$. The device has a 500μm diameter and contains more than 700 interleaved radial rows of nanofins. **b**,**c**, Transverse intensity distributions of the light emerging from the metasurface for opposite handedness of the incident light. **d**,**e**, Simulated intensity patterns for collinear 5 and 10 topological charge beams. **f**, Sketch of the setup that allows illumination of the transparent devices from the glass and air side simultaneously with circularly polarized beams of opposite handedness. This setup was also used to obtain the intensity distributions of Fig. 5 **b** and **c**.